# Applications of Various Space-time Transformations to Determine Radar Signal Distortion Caused by a Moving Target Having Constant Velocity and Acceleration

James A Boehm III[1], Muhammad Dawood[2]
1. Retired ARL engineer and Private Consultant, Las Cruces, NM, gabwolley@yahoo.com
2. Klipsch School of Electrical and Computer Engineering, New Mexico State University, Las Cruces, NM, dawood@nmsu.edu

*Abstract*— The effects of target motion on the distortion of radar signals are investigated using five transformations, namely, Hsu, Lorentz, Galilean, Reference, and Classical transformation equations. Hsu transformation is used as a primary transformation since it expresses the temporal and spatial transformations between an inertial reference frame and accelerating frame where the origin of the accelerating frame has an initial velocity and acceleration with respect to the inertial frame. Additionally, as the acceleration approach zero and infinity, respectively, the Lorentz and Galilean transformations are obtained. These transformations are used to express the transmitted waveforms in the radar reference frame variables to that of the target reference frame variables, and the reflected waveform from the target in its reference frame variable to that of the radar reference frame variables, leading to the distorted waveform received at the radar receiver due to the motion of the target.

## I. INTRODUCTION

The method (Classical) widely used today to predict the distortion due to the constant velocity and acceleration is described in [1]. In [1] the author assumes that the return signal as seen by the radar receiver is of the form $Af[t-\tau(t)]$, where $f[t]$ is the radar transmitted waveform, A is a constant, and $\tau(t)$ is the delay time for a photon to reach a point target located at range $r(t)_T$ and return to the radar receiver. The author then assumes that $\tau(t)$ can be expanded in a power series and truncated at the second order term and that this expression is accurate over the length of the radar's transmitted signal. The coefficients of the truncated approximation are then related to the velocity and acceleration of the moving point target located at range $r(t)$. Based on this assumption and given that the transmitted waveform is described by a function of *t* only, all we have to do is to substitute for *t* the new variable $t(1-2v_0/c-a_0t/c)-2x_0/c$ into *f*[*t*] to obtain the distortion of the transmitted waveform caused by the target motion.

The Classical method does not account for any relativistic effects i.e. assignment of frames of reference for the nonmoving radar transmitter and the moving target and does not establish transformation equations relating the temporal and spatial variable for each frame. Thus the expression for the returned signal does not impose any limitations on target velocity and acceleration. It also does not account for time-variant velocities and accelerations.

The Classical results have been used in [2] to show the hyperbolic frequency modulated waveform is acceleration-invariant, [3] to estimate the acceleration parameter for linear-period modulated waveform, [4] to obtain optimum frequency modulation for estimating the range, range rate and range acceleration of a moving target, and [5] to produce highly linear FM pulse radar signals.

Reference [6] addresses the distortion of the received waveform for targets having a constant velocity. All special relativity effects are included by using the Poincare transformation which includes the Lorentz transformation as a special case. References [7-13] also discuss transformations for temporal and spatial variables but consider only acceleration as a parameter in these transformations.

This paper describes a mathematical procedure to predict the distortion on a radar received waveform due to the motion of a target described by constant velocity and acceleration parameters. This method is based on the following assumptions. The first is that at the radar and target we have observers having a Cartesian reference frame and measuring rods that measure the spatial dimensions and a clock that measures the temporal dimension. The second assumption involves equations that relate the temporal and spatial variables of the observer at the radar to that of the observer located at the target and its inverse that relates the temporal and spatial variables of the observer at the target to the temporal and spatial of the observer located at the target. To this end, we consider only one spatial variable *x* and the temporal variable *t*. The general case (the three spatial variables and the single temporal variable) leads to very large mathematically intractable equations and advance nothing in determining the basic major effects of the distortion due to the target motion.

To proceed, we need a transformation relating the spatial and temporal variables that have both the velocity and acceleration of the target as parameters. References [14-19] address the issue of determining transformation of the temporal and spatial parameters as a function of velocity and accelerations. Closed forms for the transformation and its inverse are obtained. There are multiple closed forms of the transformation and its inverse based on the metric used in the frame of the acceleration observe located on the target. For this paper, we use the transformation described in [16] and call it the Hsu transformation. We use this transformation because as the acceleration approaches zero, the transformation equation turn into the Lorentz transformation equations, and as the propagation speed of the transmitted waveform approaches infinity, the transformation turns into the Galilean transformation. By a scaler transformation $w = ct$ the Hsu transformation equations can be expressed in terms of acceleration and velocity as measured by the temporal variable *t*. Additionally, close form equations are provided in [16].

In section II we show that given any transformation of the temporal and spatial variables and its inverse (in closed form) and any radar transmitted waveform prorogated in spatial dimension *x*, the distortion of the received waveform due to the target motion defined by constant velocity and acceleration can be found in closed form. Section III defines the Hsu transformation and its inverse and demonstrates how the Lorentz and Galilean transforms are obtained from it. Also defined is the Classical transformation and the Reference transformation. Section IV Defines the transmitted waveforms used as examples in the paper. Section V shows the closed form solution of each of the received waveforms assuming the Hsu, Lorentz, Galilean, Reference, and Classical transformation equations. Section VI shows numerical examples of the distortion of the target motion using the results from section V.

## II. APPLICATION OF GENERAL TRANSFORMATION EQUATIONS TO DETERMINE DISTORTION DUE TO TARGET MOTION

This section discusses mathematical operations to derive various equations. The physical meanings of these equations, variables, and parameters are discussed only when these are applied to the problem of obtaining the distortion of radar signals caused by the motion of the target.

In this analysis only the evolution variable *w* and one spatial variable *x* will be used. This is done for the following reasons. The first is that it simplifies the expressions. The second is the algebra produces very large and complex expressions even for a single spatial variable. The third is computation time is very long even for a single spatial variable. Due to the above Mathematica 10.4 [20] is used for all calculations and symbolic operations in this paper.

The equations derived here will be a function of two set of variables *x'* and *w'* and *x* and *w*, three constants $c_1, c_2, c_3$ and a function $r(c_1, c_2, c_3) = c_4$ that is a constant since it's a function of only $c_1, c_2, c_3$. The following quote from [14] defines the variable *x* and *w*, "As usual, we describe and "event" using the four coordinates $(w, x, y, z) = x^u$, and $(w', x', y', z') = x'^u$ where $u = 0, 1, 2, 3$ in inertial frames $F(w, x, y, z)$ and $F'(w', x', y', z')$ respectively. The variables *w* and *w'* are the evolution variables of a physical system (time) measured in units of length. We use the letter *w* to distinguish it from *t*, which is taken to be the evolution variable expressed using the convention unit second and also refer to *w* and *w'* as the "Taiji time.". In special relativity, the quantities *w* and *t* are related by $w = (299792458 \text{ m/s})t$ in every inertial frame. However, in Taiji relativity, since the unit of second does not exist, the variable t is not defined."

It is assumed that we have two Cartesian reference frames *F'* whose axes are labeled *x'*, *w'* and *F* whose axes are labeled *x, w*. It is also assumed that we have a relationship between the primed and unprimed coordinates

$$t' = h_1(w, x, c_1, c_2); \quad x' = h_2(w, x, c_1, c_2) \quad (1)$$

And that the inverse exists

$$t = g_1(w', x', c_1, c_2); \quad x = g_2(w', x', c_1, c_3). \quad (2)$$

In *F'* we define the scalar function

$$f_{scaler}(w', c_{ref}) \quad (3)$$

Where $c_{ref}$ is a set of constant parameters that are used in the definition of (3).

Equation (3) is propagated in the positive x' direction with a prorogation constant $c_p$

$$f_{propF'}(w', x', c_{ref}) = f_{scaler}\left(w' - \frac{x'}{c_p}, c_{ref}\right) \tag{4}$$

The prorogated waveform (4) can be expressed in terms of the w and t variables as

$$f_{propF}(w, x, c_{ref}, c_p, c_1, c_2)$$
$$= f_{scaler}\left(h_1(w, x, c_1, c_2) - \frac{h_2(w, x, c_1, c_2)}{c_p}, c_{ref}\right) \tag{5}$$

Equation (5) is the propagated function as seen by an observer located in the F frame.

Substituting the value of $c_4$ for x in (5), we have

$$f_{c4}(w, c_i) = f_{propF}(w, c_i) \tag{6}$$

Where, $c_i = c_4, c_{ref}, c_p, c_1, c_2$.

Eqn. (6) is the propagated waveform observed by the observer located in the F frame at $x = c_4$.

Eqn. (6) is now converted into a traveling wave function by propagating it in the negative x direction with the same propagation constant $c_p$

$$ff(w, x, c_i) = f_{c4}\left(w - \frac{x}{c_p}, c_i\right) \tag{7}$$

Where $c_i$ is as defined under (6).

Eqn. (7) in terms of the w' and x' variables is

$$fff(w', x', c_i) = ff\left[g_1(w', x', c_1, c_2), c_p, c_4, c_{ref}\right] \tag{8}$$

Where $c_i$ is as defined under (6).

Eqn. (8) is the prorogated waveform from the value of $x = c_4$ as seen by the observer located in the $F'$ frame, and when evaluated at x' = 0 is given as:

$$ffff(w', c_i) = fff(w', 0, c_i) \tag{9}$$

Where $c_i$ is as defined under (6).

Eqn. (9) is the prorogated waveform at x' = 0 for the observer located in the F' frame.

III. DEFINITIONS OF THE TRANSFORMATION EQUATIONS

We evaluate (9) using five transformation equations, Hsu, Lorentz, Galilean, Reference, and Classical.

In the real world we do not make measurements of the evolution variable w and w'. We use clocks to make measurements of the evolution variables. To this end, we let w = ct and w'=ct', where $c = 3 \times 10^8$. Note that $\beta$ is the rate of change of x' verses the evolutions variable w', and $\beta_0$ is the value of this rate of change evaluated at w'=0. We have that

derivative of $\beta$ is:
$$\beta = \frac{dx'}{dw'} = \frac{dw'}{dt'}\frac{dt'}{dw'} = \frac{dw'}{dt'}\frac{1}{c} = \frac{v}{c}, \text{ where } \frac{dw'}{dt'} = v \text{ the velocity in the } x' \text{ and } w' \text{ measurement frame. Note that the second}$$

$$\beta' = \frac{dx'}{dw'}\frac{dx'}{dw'} = \left(\frac{dx'}{dt'}\frac{dt'}{dw'}\right)\left(\frac{dx'}{dt'}\frac{dt'}{dw'}\right)$$

$$= \left(\frac{dx'}{dt'}\frac{dx'}{dt'}\right)\left(\frac{dt'}{dw'}\frac{dt'}{dw'}\right) = \frac{dx'^2}{dt'^2}\frac{1}{c^2} = \alpha\frac{1}{c^2}$$

Thus, the acceleration as measured by the evolution variable w' and the acceleration as measured by the evolution variable t' is scaled by the constant $1/c^2$.

We define the positive values of v and $\alpha$ so that the target will move in the positive direction in the x' frame under the action of v and $\alpha$. For this to happen, we make the substitutions w' = ct', w = ct, $\beta = -v/c$ and $\alpha$ is scaled by $-1/c^2$.

From [16] we have the Hsu transformation and its inverse expressed as a function of the constants $\alpha_0$ in $m/\sec^2$ and $v_0$ in m/sec where $c_p = c = 3 \times 10^8$:

$$t' = -\frac{v_0 F_2}{\alpha_0} + \frac{c(F_3 t + F_1)F_2^2}{\alpha_0 F} - \frac{(F_3 t + F_1)x}{cF} \tag{10a}$$

$$x' = \frac{c^2 F_2}{\alpha_0} - \frac{c^2 F_2^2}{\alpha_0 F} + \frac{x}{F}. \tag{10b}$$

Where, $F = \sqrt{1-(F_1+F_3 t)^2}$; $F_1 = \frac{v_0}{c}$; $F_2 = \sqrt{1-F_1^2}$; $F_3 = \frac{\alpha_0}{c}$.

The inverse of (10a, b) to obtain $t$ and $x$ in terms $t'$ and $x'$ is:

$$t = -\frac{cF_1 F_3^2 (x')^2 + cF_3 t'x' + c^2 F_2 t' + cF_1 F_2 x'}{F_3^2 (x')^2 + c^2 F_1^2 - c^2}; \tag{11a}$$

$$x = \left[\frac{\left(1+F_1 - \frac{\text{Num}}{\text{Den}}\right)}{\left(1-F_1 + \frac{\text{Num}}{\text{Den}}\right)}\right]^{\frac{1}{2}} \left(\frac{c^2 F_2}{\alpha_0} + x'\right) + \frac{c^2}{\alpha_0} - \frac{v_0^2}{\alpha_0} \tag{11b}$$

Where the numerator (Num) and the denominator (Den) are: $\text{Num}(t',x') = \alpha_0 \left(F_1 F_3 x'^2 + \alpha_0 t'x' + c^2 F_2 t' + v_0 F_2 x'\right)$ and $\text{Den} = cF_3^2 (x')^2 + c^3 F_1^2 - c^3$.

Although, the inverse was obtained using Mathematica, we believe it is the same inverse as shown in [16]. Further, as the acceleration approaches zero we obtain the inverse Lorentz transformations while the inverse in [16] does not.

Using Mathematica's Limit function we can show that as $\alpha_0 \to 0$ (10 a, b) reduce to

$$\lim_{a_0 \to 0} t' = \frac{t}{F_2} - \frac{v_0 x}{c^2 F_2}; \quad \lim_{a_0 \to 0} x' = -\frac{tv_0}{F_2} + \frac{x}{F_2} \tag{12}$$

And (11 a, b) become:

$$\lim_{a_0 \to 0} t = \frac{t'}{F_2} + \frac{v_0 x'}{c^2 F_2}; \quad \lim_{a_0 \to 0} x = \frac{t'v_0}{F_2} + \frac{x'}{F_2} \tag{13}$$

Which are the Lorentz transformations.

If we assume the principle of simultaneity i.e. that speed of the prorogation $c \to \infty$ then (10) and (11) reduce to the Galilean transformation.

$$t' = t; \quad x' = -\frac{\alpha_0 t^2}{2} - t v_0 + x \tag{14}$$

And

$$t = t'; \quad x = \frac{\alpha_0 t^2}{2} + t v_0 + x' \tag{15}$$

The reference transformation is: $t = t'; \quad x = x'$.

The classical transformation in [1] is obtained by scaling $t$ by the factor $t = t\left(1 - \frac{2v_0}{c} - \frac{\alpha_0 t}{c}\right) - \frac{2x_0}{c}$ in the prorogated waveform to obtain the waveform received by the radar.

## IV. RADAR WAVEFORMS DEFINED

We analysis the distortion effects of constant velocity and acceleration for five reference waveforms that are transmitted by the radar at the carrier frequency $f_c$ with a pulse width of $pw$.

The first is the pulsed sinewave

$$\text{sinwave}_{ref}(\cdot) = \exp\left[i2\pi f_c t \{\Phi(t) - \Phi(t - pw)\}\right]$$

The second is the chirp waveform

$$\text{chirp}_{ref}(\cdot) = \exp\left[-i2\pi\left(f_c t + \text{slope} \cdot t^2\right)\right]\{\Phi(t) - \Phi(t - pw)\}$$

The third is the hyperbolic waveform

$$\text{hyperbolic}_{ref}(\cdot) = \exp\left[\frac{-i2\pi}{b}\log\left(1 + b f_1 t\right)\right]\{\Phi(t) - \Phi(t - pw)\}$$

The fourth is the thirteen bit barker code

$$\text{barker13}_{ref}(\cdot) = \sum_{i=1}^{13} \exp\left[-i2\pi f_c t\right] BC_i \{\Phi(..) - \Phi(.)\}$$

Where, $BC = \{1,1,1,1,1,-1,-1,1,1,-1,1,-1,1\}$ and $\Phi(..) = \Phi\left(t - \left(i - 1\frac{pw}{13}\right)\right)$ $\Phi(.) = \Phi\left(t - i\frac{pw}{13}\right)$.

The fifth is a carrier AM modulated by a sequence of 13 pulses whose amplitudes are generated by selections from a normal Gaussian random variable

$$\text{gaussian}_{ref}(\cdot) = \sum_{i=1}^{32} \exp\left[-i2\pi f_c t\right] \text{Gaussian}_i \{\Phi(..) - \Phi(.)\}$$

For all waveforms $\Phi(t)$ is the Heaviside step function.

## V. WAVEFORMS EXPRESSIONS AS SEEN BY THE RADAR

The expressions for each of the waveforms seen by the radar receiver are very large. The expression for the received chirp waveform converted to a pdf format would be approximately twenty pages long. The time to calculated 3000 samples of the received chirp waveform using Mathematica 10.4 on a standard laptop is in the order of minutes. So something is needed to increase the speed of computation and to display the expressions in a compact form for the various transformations acting on the five transmitted waveforms. Each transformed waveform was looked at for repeated expressions involving the various constants. These expressions can be defined in the each waveform case to simplify the expression and by pre calculating them before calculating the expression the calculations can be made faster. Using the selected expressions results can be displayed in a more compact manner and calculation speed increased 10 to 20 fold.

The expressions needed to reduce the received waveforms when subjected to the Hsu transformation are:

$$x_{xintav} = -\frac{c\left(-c + v_0 + \sqrt{c^2 - 2cv_0 + v_0^2 - 2\alpha_0 x_0}\right)}{\alpha_0};$$

$$F_{1W} = F_2;\ F_{2W} = -F_{1W} F_1 x_{xintav};\ F_{3W} = -\frac{F_1 F_3 x_{xintav}^2}{c};$$

$$F_{4W} = -F_2^2 + \frac{F_3^2 x_{xintav}^2}{c^2};\ F_{5W} = F_2^2;\ F_{6W} = -\frac{F_{1W} v_0}{F_3};$$

$$F_{7W} = \frac{cF_{1W}}{F_3};\ F_{8W} = F_1;\ F_{9W} = c^2 F_{5W};$$

$$F_{10W} = F_{7W} - x_{xintav};\ F_{11W} = \alpha_0 x_{xintav}.$$

In addition, the following functions for $t'$ are needed:

$$F_{12W} = F_{7W} - \frac{F_{9W}}{\alpha_0 \sqrt{1-(F_{8W}+F_3 t')^2}};$$

$$F_{13W} = -F_{8W} - \frac{F_3}{c}\left(F_{6W} - \frac{c^2 F_{5W}(F_{8W}+F_3 t')}{\alpha_0 \sqrt{1-(F_{8W}+F_3 t')^2}}\right);$$

$$F_{14W} = -F_{8W} - \frac{F_3}{c}\left(F_{6W} - \frac{c^2 F_{9W}(F_{8W}+F_3 t')}{\alpha_0 \sqrt{1-(F_{8W}+F_3 t')^2}}\right);$$

$$F_{15W} = \frac{1}{c}\left(F_{10W} + \frac{F_{12W}}{\sqrt{1-F_{14W}^2}} - \frac{F_{9W}}{\alpha_0\sqrt{1-F_{13W}^2}} + \frac{F_{12W} F_{14W}}{\sqrt{1-F_{14W}^2}} + F_{6W} - \frac{F_{13W} F_{19W}}{\alpha_0\sqrt{1-F_{13W}^2}}\right);$$

$$F_{16W} = \frac{F_3}{cF_{4W}}\left(-\frac{F_{11W} F_{15W}}{c} - cF_{15W} F_{1W} + F_{2W} + F_{3W}\right);$$

$$F_{17W} = \frac{1}{cF_{4W}}\left(-\frac{F_{11W} F_{15W}}{c} - cF_{15W} F_{1W} + F_{2W} + F_{3W}\right).$$

The expressions needed to reduce the received waveform subjected to the Lorentz transformation are (note we do not need expressions involving $t$):

$$x_{intxv} = \frac{cx_0}{c-v_0}; F_{1L} = F_2; F_{2L} = \frac{F_{1L}v_0 x_{intxv}}{c^2 - v_0^2}$$

$$F_{3L} = \frac{c^2 F_{1L} x_{intx}}{c^2 - v_0^2}; F_{4L} = c^2 - v_0^2;$$

$$F_{5L} = c^2 F_{1L} v_0; F_{6L} = c^2 F_{1L}; F_{7L} = F_{1L}^2$$

$$F_{8L} = F_1^2; F_{9L} = \frac{F_{8L}}{F_{7L}}.$$

We need to define other transformation equations in order to further simplify the displayed expressions for the received waveform for each reference under the Hsu and Lorentz transformations. These transformation equations were not used in any of the numerical calculations.

For the Wu transformation we need the following

$$F_{18W} = \sqrt{(1 + F_{16W} + F_{8W})(1 - F_{16W} - F_{8W})}$$

$$F_{19W} = \sqrt{(+1 + F_{8W})(1 - F_{8W})}$$

And for the Lorentz transformation we need the following

$$F_{10L} = \frac{t'}{F_{17L}} + F_{19L} t' - \frac{1}{c}\left(x_{xintv} + \frac{2v_0 t'}{F_{17L}}\right)$$

Appendices A-D show the expressions for the received waveforms under the various transformations used in this paper. Numerical Analysis

A. *General*

Figures 1-5 show the velocity- and acceleration-dependent distortion effects on the received waveforms for various transformations. Each figure has a reference transformation which is the identity transformation of the spatial and temporal variables. This reference is used to compare the effects of the velocity and accelerations of the target for each transformation, namely, Hsu, Lorentz, Galilean, and Classical. As mentioned previously, the Classical transformation is used extensively to predict the effects of target motion on the received waveform. Therefore, Classical transformation is also included in each figure to compare and contrast with the results pertaining to the other transformations.

Five waveforms were selected for graphical analysis. They are pulsed sinewave, chirp pulse, hyperbolic chirp pulse, 13 bit barker code pulse, and 13 bit random amplitude pulse. In order to compare the distortion effects across waveforms the following parameters were used for each waveform.

carrier frequency = $3 \times 10^8$ Hz;

pulse width = .0001 sec;

sample time = $1.32118 \times 10^{-9}$ sec;

initial range to the target = 6000.18 m;

constant velocity = 15725.0 m/sec;

constant acceleration = $2. \times 10^8$;

simulation time = .00029 sec;

number of samples = 219507;

v0 = 15625 m /sec; a0 = $2 \times 10^8$ m/sec$^2$.

In order to compare the chirp and hyperbolic waveforms, we adjusted the *slope* and *b* values so that the start and stop frequencies were the same for each waveform. To compare the 13 bit barker coded waveform with the random amplitude sub pulses we used 13 sub pulses.

In all cases, if the amplitude and spectrum are compared to the reference's amplitude and spectrum no effects of the targets motion can be observed. In order to observe the effects of the target motion the output of the matched filter created by correlating the radar observed signal for the Hsu, Lorentz, Galilean, and Classical transformations with that of the observed signal for the reference waveform is used. For the pulsed sinewave the matched filter used is the spectrum of the received signal.

### B. PULSED SINEWAVE ANALYSIS

In Figure 1, we see the expected basic results under no acceleration. The matched filter output is merely shifted in frequency with respect to the reference. The shift in frequency for the Hsu, Lorentz, and Galilean transformations are the same and are greater than that for the Classical case. The effects of the acceleration are as expected i.e. a broadening of the spectrum with the broadening being equal between Hsu and Galilean transformations, and greater than the Classical case. Lorentz transformation does not show the effects of acceleration.

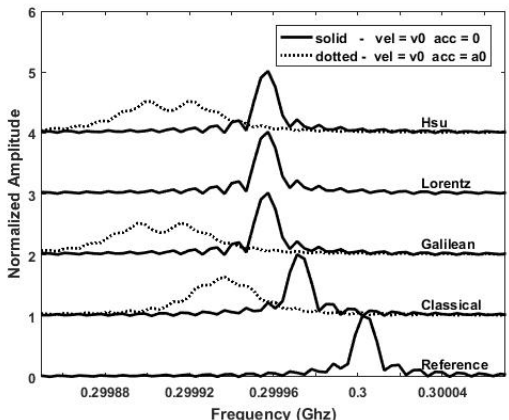

Figure 1: Received Pulsed Sinewave Waveforms for each Transformation

### C. CHIRP ANALYSIS

The specific parameters used for the chirp analysis are: Chirp start frequency: $3 \times 10^8$, and chirp bandwidth: $1.5 \times 10^8$. These parameters produced a chirp stop frequency of $4.5 \times 10^8$ and a time-bandwidth product = 15000. In Figure 2, we observe under zero acceleration that the matched-filter output is not only shifted in time, it is distorted as well. The distortion is more pronounced for Hsu, Lorentz, and Galilean transformations than the Classical case. When acceleration is introduced, a broadening effect is observed for all transformations except Lorentz which is invariant to the acceleration.

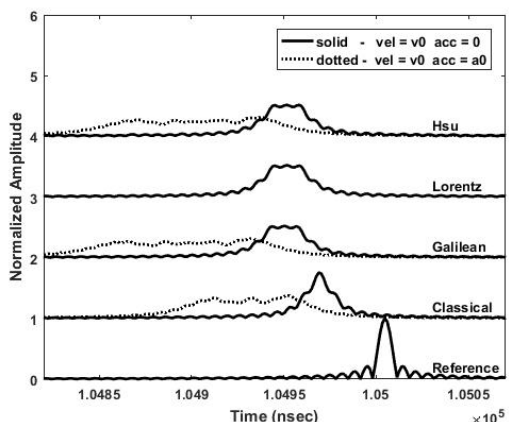

Figure 2: Received Chirp Waveforms for each Transformation

### A. HYPERBOLIC ANALYSIS

The specific parameters used in the hyperbolic analysis are: start and stop frequency of $3 \times 10^8$, and $4.5 \times 10^8$, respectively. These parameters produce the same stop frequency and time bandwidth product as the chirp waveform with the parameter $b = -.0000111108$.

The hyperbolic waveform does not show that acceleration has no effect on the received waveform as predicated by [2]. This reference assumes that the signal is a baseband but our waveform has a carrier frequency. It is this difference that causes the distortion of the received waveform when acceleration is considered. The remarkable result is that when compared to the chirp waveform having the same spectrum and time bandwidth product, the velocity produces no distortion and only creates a time shift in the matched filter output when compared to the reference waveform's compressed output. Lorentz transformation is acceleration-invariant.

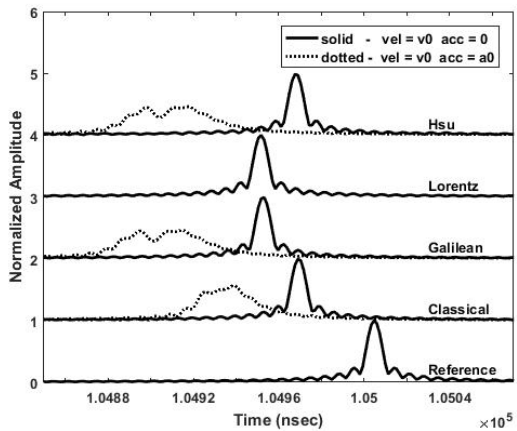

Figure 3: Received Hyperbolic Waveforms for each Transformation

### B. BARKER CODE ANALYSIS

The 13 bit barker code sequence is used for this simulation. In Figure 4, we observe that the output of the matched filter shows no compression effects at all even under zero acceleration condition. Thus, Barker coded waveforms are very sensitive to velocity as well as to the acceleration.

### C. GAUSSIAN ANALYSIS

For this case, a 13 bit random sequence drawn from a standard Gaussian probability density function is used to generate 13 sub pulses. Again we observe in Figure 5 that even with zero acceleration the output of the matched filter shows no compression effects at all.

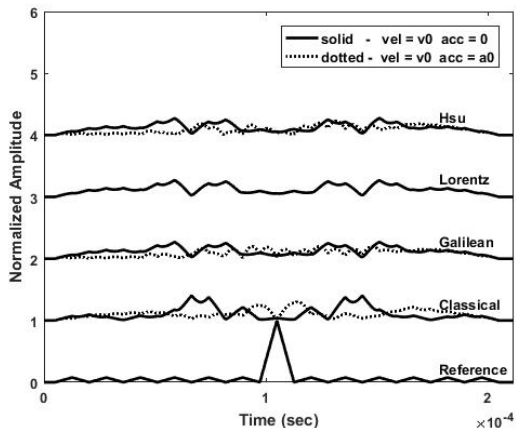

Figure 4: Received Barker Code Waveforms for each Transformation

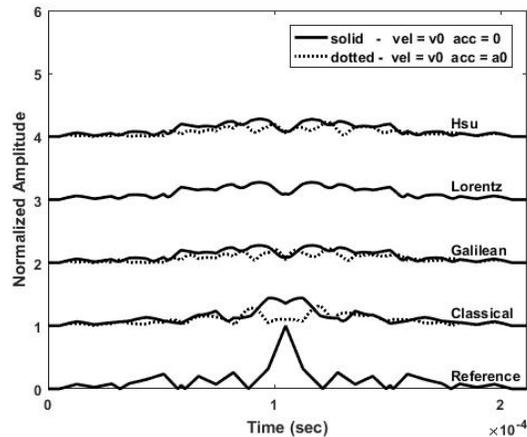
Figure 5: Received Gaussian Amplitude Waveforms for each Transformation

VI. CONCLUSIONS

It is shown in this paper that the distortion effects due to target velocity and acceleration on the received waveform are in general agreement with the standard Classical transformation. Thus, the Classical transformation is a good indicator of the general effects of target motion on the distortion of the received waveform for all five waveforms considered in this paper. The Hsu transformation, however, predicated larger distortions with targets having acceleration components. Of all the waveforms considered, the pulse hyperbolic waveform had less distortion effects due to target motion. The pulsed Barker coded and Gaussian amplitude waveforms were not effective at all when the target had motion.

We also considered the distortion effects when the velocity and acceleration were kept constant and the time bandwidth of the waveform was varied, showing that as the time bandwidth of the signal increased so did the distortion, for all waveform considered. We also considered different velocity and acceleration scenarios while keeping the time-bandwidth product constant. These analysis indicate that as the velocity and acceleration decreased the distortion effects approached the effects predicated by the Classical transformation, and, in addition, as both the velocity and acceleration approached zero the distortion approached that of the reference transformation. These results held for all waveforms considered. These analysis are however are not shown in this paper due to length constraints.

# Appendices
## Appendix A: Pulsed Sin Wave

The expression for the pulsed sinewave that is seen by the radar receiver using the Hsu transformation is:

$$f_{recSINW} = \exp\left\{2i\pi f_c \left(F_{17W} - \frac{1}{c}\left(\frac{c^2}{\alpha_0} - \frac{c^2 F_{19W} F_{18W}}{\alpha_0} - \frac{v_0^2}{\alpha_0} - F_{18W} x_{xintav}\right)\right)\right\}$$

$$\left(\Phi\left(F_{17W} - \frac{1}{c}\left(\frac{c^2}{\alpha_0} - \frac{c^2 F_{19W} F_{18W}}{\alpha_0} - \frac{v_0^2}{\alpha_0} - F_{18W} x_{xintav}\right)\right) - \Phi\left(F_{17W} - \frac{1}{c}\left(\frac{c^2}{\alpha_0} - \frac{c^2 F_{19W} F_{18W}}{\alpha_0} - \frac{v_0^2}{\alpha_0} - F_{18W} x_{xintav}\right) - pw\right)\right)$$

The expression of the pulsed sinewave as seen by the radar receiver via the Lorentz transformation is:

$$f_{recSINL} = \exp\left\{2i\pi f_c \left[F_{2L} + \frac{F_{6L} F_{10L}}{F_{4L}} - \frac{1}{c}\left(F_{3l} + \frac{F_{5L} F_{10L}}{F_{4L}}\right)\right]\right\}$$

$$\left[\Phi\left(F_{2L} + \frac{F_{6L} F_{10L}}{F_{4L}} - \frac{1}{c}\left(F_{3l} + \frac{F_{5L} F_{10L}}{F_{4L}}\right)\right) - \Phi\left(F_{2L} + \frac{F_{6L} F_{10L}}{F_{4L}} - \frac{1}{c}\left(F_{3l} + \frac{F_{5L} F_{10L}}{F_{4L}}\right) - pw\right)\right]$$

The expression for the pulsed sinewave as seen by the radar receiver via the Galilean transformation is:

$$frec\sin G = \exp\left[-\frac{2i\pi f_c}{c}\begin{pmatrix}ct + -\alpha t^2 - 2tv_0 - x\text{intav} \\ -v_0\left(t - \frac{\alpha t^2 + 2tv_0 + x\text{intav}}{c}\right) + \frac{\alpha_0}{2}\left(t - \frac{\alpha t^2 + 2tv_0 + x\text{intav}}{c}\right)^2 + x\text{intav}\end{pmatrix}\right]$$

$$\left(\Phi\left[\frac{1}{c}\left\{(ct - \alpha t^2 - 2tv_0 - x\text{intav}) - v_0\left(t - \frac{\alpha t^2 + 2tv_0 + x\text{intav}}{c}\right) + \frac{\alpha_0}{2}\left(t - \frac{\alpha t^2 + 2tv_0 + x\text{intav}}{c}\right)^2 + x\text{intav}\right\}\right]\right.$$

$$\left. - \Phi\left[\frac{1}{c}\left\{(ct - \alpha t^2 - 2tv_0 - x\text{intav}) - v_0\left(t - \frac{\alpha t^2 + 2tv_0 + x\text{intav}}{c}\right) + \frac{\alpha_0}{2}\left(t - \frac{\alpha t^2 + 2tv_0 + x\text{intav}}{c}\right)^2 + x\text{intav}\right\} - pw\right]\right)$$

The expression for the pulsed sinewave as seen by the radar receiver via the Classical transformation is:

$$f_{rec\sin c} = \sin\left[\frac{2\pi f_c}{c}\left\{t(c - \alpha_0 t' - 2x_0) - 2x_0\right\}\right]$$

$$\left[\Phi\left(\frac{1}{c}\left\{t(c - \alpha_0 t' - 2x_0) - 2x_0\right\}\right) - \Phi\left(\frac{1}{c}\left\{t(c - \alpha_0 t' - 2x_0) - 2x_0\right\} - pw\right)\right]$$

## Appendix B: Chirp Waveform

The expression for Chirp waveforms as seen by the radar receiver under different transformations are as follows:

### B1: Hsu transformation

$$f_{\text{recCHIRPW}} =$$

$$\exp\left\{-2i\pi\left[f_c\left(\frac{F_{17W}}{c}\left\{\frac{c^2}{\alpha_0}-\frac{c^2 F_{19W} F_{18W}}{\alpha_0}-\frac{v_0^2}{\alpha_0}+F_{18W} x_{\text{xintav}}\right\}\right)+slope\left(\frac{F_{17W}}{c}\left\{\frac{c^2}{\alpha_0}-\frac{c^2 F_{19W} F_{18W}}{\alpha_0}-\frac{v_0^2}{\alpha_0}+F_{18W} x_{\text{xintav}}\right\}\right)^2\right]\right\}$$

$$\left(\Phi\left(\frac{F_{17W}}{c}\left(\frac{c^2}{\alpha_0}-\frac{c^2 F_{19W} F_{18W}}{\alpha_0}-\frac{v_0^2}{\alpha_0}+F_{18W} x_{\text{xintav}}\right)\right)-\Phi\left(\frac{F_{17W}}{c}\left(\frac{c^2}{\alpha_0}-\frac{c^2 F_{19W} F_{18W}}{\alpha_0}-\frac{v_0^2}{\alpha_0}+F_{18W} x_{\text{xintav}}\right)-pw\right)\right)$$

The expression for the chirp waveform that is seen by the radar receiver using the Lorentz transformation is:

$$f_{\text{recCHIRPL}} = \exp\left\{-2i\pi\left[f_c\left(F_{2L}+\frac{F_{6L}F_{10L}}{F_{4L}}-\frac{1}{c}\left(F_{3L}+\frac{F_{5L}F_{10L}}{F_{4L}}\right)\right)+slope\left(F_{2L}+\frac{F_{6L}F_{10L}}{F_{4L}}-\frac{1}{c}\left(F_{3L}+\frac{F_{5L}F_{10L}}{F_{4L}}\right)\right)^2\right]\right\}$$

$$\left(\Phi\left(\frac{1}{c}\left(F_{3L}+\frac{F_{5L}F_{10L}}{F_{4L}}\right)\right)-\Phi\left(\frac{1}{c}\left(F_{3L}+\frac{F_{5L}F_{10L}}{F_{4L}}\right)-pw\right)\right)$$

### B2: Galilean transformation

$$freechirpG = \exp\left[-2i\pi\left[\begin{array}{c}\frac{f_c}{c}\left(\begin{array}{c}ct-\alpha t^2-2tv_0-\text{xintav}\\-v_0\left(t+\frac{-\alpha t^2-2tv_0-\text{xintav}}{c}\right)-\frac{\alpha_0}{2}\left(t+\frac{-\alpha t^2-2tv_0-\text{xintav}}{c}\right)^2+\text{xintav}\end{array}\right)\\+slope\left[\frac{1}{c}\left(\begin{array}{c}ct-\alpha t^2-2tv_0-\text{xintav}\\-v_0\left(t+\frac{-\alpha t^2-2tv_0-\text{xintav}}{c}\right)-\frac{\alpha_0}{2}\left(t+\frac{-\alpha t^2-2tv_0-\text{xintav}}{c}\right)^2+\text{xintav}\end{array}\right)\right]^2\end{array}\right]\right]$$

$$\left(\Phi\left(\frac{1}{c}\left\{\begin{array}{c}ct-\alpha t^2-2tv_0-\text{xintav}\\-v_0\left(t+\frac{-\alpha t^2-2tv_0-\text{xintav}}{c}\right)-\frac{\alpha_0}{2}\left(t+\frac{-\alpha t^2-2tv_0-\text{xintav}}{c}\right)^2+\text{xintav}\end{array}\right\}\right)-\Phi\left(\frac{1}{c}\left\{\begin{array}{c}ct-\alpha t^2-2tv_0-\text{xintav}\\-v_0\left(t+\frac{-\alpha t^2-2tv_0-\text{xintav}}{c}\right)-\frac{\alpha_0}{2}\left(t+\frac{-\alpha t^2-2tv_0-\text{xintav}}{c}\right)^2+\text{xintav}\end{array}\right\}-pw\right)\right)$$

### B3: Classical transformation

$$f_{recchirpC} = Exp\left[-2i\pi\left\{f_c\left(t\left(1-\frac{\alpha_0 t}{c}-\frac{2v_0}{c}\right)-\frac{2x_0}{c}\right)+slope\left(t\left(1-\frac{\alpha_0 t}{c}-\frac{2v_0}{c}\right)-\frac{2x_0}{c}\right)^2\right\}\right]$$

$$\left[\Phi\left(\left(t\left(1-\frac{\alpha_0 t}{c}-\frac{2v_0}{c}\right)-\frac{2x_0}{c}\right)\right)-\Phi\left(\left(t\left(1-\frac{\alpha_0 t}{c}-\frac{2v_0}{c}\right)-\frac{2x_0}{c}\right)-pw\right)\right]$$

# Appendix C: Hyperbolic FM Waveform

The expression for hyperbolic waveform as seen by the radar receiver under different transformations are as follows:

## C1: Hsu transformation

$f_{\text{recHYPW}} =$

$$\left(1 - bf_1\left(\frac{F_{17W}}{c}\left\{\frac{c^2}{\alpha_0} - \frac{c^2 F_{19W} F_{18W}}{\alpha_0} - \frac{v_0^2}{\alpha_0} + F_{18W} x_{\text{xintav}}\right\}\right)\right)^{-\frac{2i\pi}{b}}$$

$$\left(\Phi\left(\frac{F_{17W}}{c}\left\{\frac{c^2}{\alpha_0} - \frac{c^2 F_{19W} F_{18W}}{\alpha_0} - \frac{v_0^2}{\alpha_0} + F_{18W} x_{\text{xintav}}\right\}\right) - \Phi\left(\frac{F_{17W}}{c}\left\{\frac{c^2}{\alpha_0} - \frac{c^2 F_{19W} F_{18W}}{\alpha_0} - \frac{v_0^2}{\alpha_0} + F_{18W} x_{\text{xintav}}\right\} - pw\right)\right)$$

## C2: Lorentz transformation

$$f_{\text{recHYPL}} = \left(1 + bf_1\left(F_{2L} + \frac{F_{6L} F_{10L}}{F_{4L}} - \frac{1}{c}\left(F_{3L} + \frac{F_{5L} F_{10L}}{F_{4L}}\right)\right)\right)^{-\frac{2i\pi}{b}}$$

$$\left(\Phi\left(F_{2L} + \frac{F_{6L} F_{10L}}{F_{4L}} - \frac{1}{c}\left(F_{3L} + \frac{F_{5L} F_{10L}}{F_{4L}}\right)\right) - \Phi\left(F_{2L} + \frac{F_{6L} F_{10L}}{F_{4L}} - \frac{1}{c}\left(F_{3L} + \frac{F_{5L} F_{10L}}{F_{4L}}\right) - pw\right)\right)$$

## C3: Galilean transformation

$$rechypG = \left(1 + bf_1\left(\frac{1}{c}\left\{\begin{array}{l}ct - \alpha t^2 - 2tv_0 - x\text{intav} \\ -v_0\left(t + \frac{-\alpha t^2 - 2tv_0 - x\text{intav}}{c}\right) - \frac{\alpha_0}{2}\left(t + \frac{-\alpha t^2 - 2tv_0 - x\text{intav}}{c}\right)^2 + x\text{intav}\end{array}\right\}\right)\right)$$

$$\left(\Phi\left(\frac{1}{c}\left\{\begin{array}{l}ct - \alpha t^2 - 2tv_0 - x\text{intav} \\ -v_0\left(t + \frac{-\alpha t^2 - 2tv_0 - x\text{intav}}{c}\right) - \frac{\alpha_0}{2}\left(t + \frac{-\alpha t^2 - 2tv_0 - x\text{intav}}{c}\right)^2 + x\text{intav}\end{array}\right\}\right)\right.$$

$$\left. -\Phi\left(\frac{1}{c}\left\{\begin{array}{l}ct - \alpha t^2 - 2tv_0 - x\text{intav} \\ -v_0\left(t + \frac{-\alpha t^2 - 2tv_0 - x\text{intav}}{c}\right) - \frac{\alpha_0}{2}\left(t + \frac{-\alpha t^2 - 2tv_0 - x\text{intav}}{c}\right)^2 + x\text{intav}\end{array}\right\} - pw\right)\right)$$

## C4: Classical transformation

$$f_{\text{rechypC}} = \left(1 + b f_1\left(t'\left(1 - \frac{\alpha_0 t'}{c} - \frac{2v_0}{c}\right) - \frac{2x_0}{c}\right)\right)^{-\frac{2i\pi}{b}}$$

$$\left(\Phi\left(t'\left(1 - \frac{\alpha_0 t'}{c} - \frac{2v_0}{c}\right) - \frac{2x_0}{c}\right) - \Phi\left(t'\left(1 - \frac{\alpha_0 t'}{c} - \frac{2v_0}{c}\right) - \frac{2x_0}{c} - pw\right)\right)$$

## Appendix D: Barker and Gaussian Codes

The expressions for the Barker and Gaussian codes have the same form. The only difference is the value of the code at each index i and the number of pulses N summed over. The expression for the coded waveform as seen by the radar receiver under different transformations are as follows:

### D1: Hsu transformation

$$fcodeW = \sum_{i=1}^{N} \left( \begin{array}{l} C_i \exp\left\{ 2i\pi f_c \left( F_{17W} - \dfrac{\dfrac{c^2}{\alpha_0} - \dfrac{c^2 F_{19W} F_{18W}}{\alpha_0} - \dfrac{v_0^2}{\alpha_0} - F_{18W} x_{xintav}}{c} \right) \right\} \\ \Phi\left( -(i-1) dtseg + F_{17W} - \dfrac{\dfrac{c^2}{\alpha_0} - \dfrac{c^2 F_{19W} F_{18W}}{\alpha_0} - \dfrac{v_0^2}{\alpha_0} - F_{18W} x_{xintav}}{c} \right) \\ -\Phi\left( -i\, dtseg + F_{17W} - \dfrac{\dfrac{c^2}{\alpha_0} - \dfrac{c^2 F_{19W} F_{18W}}{\alpha_0} - \dfrac{v_0^2}{\alpha_0} - F_{18W} x_{xintav}}{c} - pw \right) \end{array} \right)$$

### D2: Lorentz transformation

$$fcodeL = \sum_{i=1}^{N} \left[ \begin{array}{l} C_i \exp\left\{ 2i\pi f_c \left[ F_{2L} + \dfrac{F_{6L} F_{10L}}{F_{4L}} - \dfrac{1}{c}\left( F_{3l} + \dfrac{F_{5L} F_{10L}}{F_{4L}} \right) \right] \right\} \\ \left[ \Phi\left( -(i-1)\, dtseg + F_{2L} + \dfrac{F_{6L} F_{10L}}{F_{4L}} - \dfrac{1}{c}\left( F_{3l} + \dfrac{F_{5L} F_{10L}}{F_{4L}} \right) \right) \right. \\ \left. -\Phi\left( -i\, dtseg + F_{2L} + \dfrac{F_{6L} F_{10L}}{F_{4L}} - \dfrac{1}{c}\left( F_{3l} + \dfrac{F_{5L} F_{10L}}{F_{4L}} \right) - pw \right) \right] \end{array} \right]$$

### D3: Galilean transformation

$$f_{reccodeG} = \sum_{i=1}^{13} \left( \begin{array}{l} b_i \exp\left[ -2i\pi f_c \left( \dfrac{F_2}{c} + t' - \dfrac{1}{c}\left( F_1 + v_0\left(\dfrac{F_2}{c} + t'\right) + \dfrac{1}{2}\alpha_0 \left(\dfrac{F_2}{c} + t'\right)^2 \right) \right) \right] \\ \left( \Phi\left( -(i-1)\, dtseg + \dfrac{F_2}{c} + t' - \dfrac{1}{c}\left( F_1 + v_0\left(\dfrac{F_2}{c} + t'\right) + \dfrac{1}{2}\alpha_0 \left(\dfrac{F_2}{c} + t'\right)^2 \right) \right) \right. \\ \left. -\Phi\left( -i\, dtseg + \dfrac{F_2}{c} + t' - \dfrac{1}{c}\left( F_1 + v_0\left(\dfrac{F_2}{c} + t'\right) + \dfrac{1}{2}\alpha_0 \left(\dfrac{F_2}{c} + t'\right)^2 \right) \right) \right) \end{array} \right)$$

## D4: Classical transformation

$$f_{reccodeC} = \sum_{1}^{13} \left[ \begin{array}{c} b_i \exp[-2if_c\pi\left(t\left(1-\frac{\alpha_0 t'}{c}-\frac{2v_0}{c}\right)-\frac{2x_0}{c}\right)] \\ \left(\Phi\left(-(i-1)dtseg + t\left(1-\frac{\alpha_0 t'}{c}-\frac{2v_0}{c}\right)-\frac{2x_0}{c}\right) - \Phi\left(-i\,dtseg + t\left(1-\frac{\alpha_0 t'}{c}-\frac{2v_0}{c}\right)-\frac{2x_0}{c}\right)\right) \end{array} \right]$$

Where N is the length of the code and $dtseg = \frac{pulsewidth}{N}$ .